\documentclass[pra,aps,twocolumn, float]{revtex4-2}
\usepackage{epsfig}
\usepackage{mathrsfs}
\usepackage{mathptmx}
\usepackage{float}
\usepackage{latexsym,amssymb,amsthm,amsmath}
\usepackage{bm}
\usepackage{times}
\usepackage[colorlinks,linkcolor=blue,anchorcolor=blue,citecolor=blue,urlcolor=blue]{hyperref}
\UseRawInputEncoding
\begin{document}
\title{Enhanced spin-mechanical interaction with levitated micromagnets}
\author{Xue-Feng Pan}
\author{Xin-Lei Hei}
\author{Xing-Liang Dong}
\author{Jia-Qiang Chen}
\author{Cai-Peng Shen}
\author{Hamad Ali}
\author{Peng-Bo Li}
\email{lipengbo@mail.xjtu.edu.cn}
\affiliation{Ministry of Education Key Laboratory for Nonequilibrium Synthesis and Modulation of Condensed Matter, Shaanxi Province Key Laboratory of Quantum Information and Quantum Optoelectronic Devices, School of Physics, Xi'an Jiaotong University, Xi'an 710049, China}
\date{\today}
\begin{abstract}
% Because of the great sensitivity, low damping, and low noise in an ultra-vacuum environment, levitated oscillators have been widely employed in ultra-sensitive sensing, non-classical state preparation, ground-state cooling, and other quantum information processes.
Spin-mechanical hybrid systems have been widely used in quantum information processing. However, the spin-mechanical interaction is generally weak, making it a critical challenge to enhance the spin-mechanical interaction into the strong coupling or even ultra-strong coupling regime. Here, we propose a protocol that can significantly enhance the spin-mechanical coupling strength with a diamond spin vacancy and  a levitated micromagnet. A driving electrical current is used to modulate the mechanical motion of the levitated micromagnet, which induces a two-phonon drive and can  exponentially enhance the spin-phonon and phonon-medicated spin-spin coupling strengths. Furthermore, a high fidelity Schr\"odinger cat state and an unconventional 2-qubit geometric phase gate with high fidelity and faster gate speed can be achieved using this hybrid system. This protocol provides a promising platform  for quantum information processing with NV spins coupled to levitated micromagnets.
\end{abstract}

\maketitle

\section{\label{sec:I}introduction}
Hybrid quantum systems, which combine the advantages of various quantum systems to overcome their shortcomings, have been widely used in quantum information processing~\cite{2013XiangP623653, 2020ClerkP257267, 2009WallquistP1400114001}. Several proposals for hybrid systems in cavity-QED~\cite{2006WaltherP13251382}, circuit-QED~\cite{2021BlaisP2500525005}, and spin-mechanical hybrid systems~\cite{2020HuilleryP134415134415, 2021DongP203601203601, 2011ArcizetP879883, 2012HongP39203924, 2012KolkowitzP16031606, 2004WilsonRaeP7550775507, 2015LiP4400344003, 2013BennettP156402156402, 2013KepesidisP6410564105, 2021StreltsovP193602193602, 2009RablP4130241302, 2020LiP153602153602,  2009XuP2233522335, 2013ChotorlishviliP8520185201, 2010ZhouP4232342323, 2010RablP602608, 2014OvartchaiyapongP44294429, 2014TeissierP2050320503, 2014AsjadP4550245502, 2018SanchezMunozP123604123604, 2013MacQuarrieP227602227602, 2015PigeauP86038603, 2018CarterP246801246801, 2016LiP1550215502} have already been implemented in recent years. The spin-mechanical hybrid system combines quantum systems with long coherence time, such as trapped atoms or ions~\cite{2013LemmerP8300183001, 2004PorrasP207901207901, 2012BrittonP489492, 2020MartinetzP101101, 2017DelordP6381063810}, solid-state spins~\cite{2020BarryP1500415004, 2013DohertyP145, 2019BradacP56255625, 2018MeesalaP205444205444, 2018LemondeP213603213603, 2014HeppP3640536405, 2021ChenP1370913709}, and mechanical oscillators with high-quality factors, such as cantilevers~\cite{2020LiP153602153602, 2009RablP4130241302, 2014TeissierP2050320503, 2014OvartchaiyapongP44294429, 2010RablP602608, 2009XuP2233522335, 2010ZhouP4232342323, 2013ChotorlishviliP8520185201} and nanobeams~\cite{2013KepesidisP6410564105, 2015LiP4400344003, 2013BennettP156402156402, 2004WilsonRaeP7550775507}. It has been widely used in the preparation of a non-classical quantum state of mechanical motion~\cite{2018SanchezMunozP123604123604, 2014AsjadP4550245502}, ground-state cooling~\cite{2009RablP4130241302, 2013KepesidisP6410564105, 2021StreltsovP193602193602}, ultrasensitive sensing~\cite{2012KolkowitzP16031606, 2017PratCampsP3400234002}, as well as the generation of interaction between two distant quantum systems~\cite{2010ZhouP4232342323, 2009XuP2233522335, 2013ChotorlishviliP8520185201, 2020LiP153602153602}. The greatest impediment to its possible applications is the unavoidable dissipation of the oscillators interacting with the environment. To reduce dissipation, researchers have developed levitated devices~\cite{2019TimberlakeP224101224101,2020VinanteP6402764027,2018SlezakP6302863028,2021NavauP174436174436,2017RusconiP134419134419, 2016JohnssonP3749537495,2019WalkerP6381963819,2021LengP2406124061,2020GieselerP163604163604,2021XiongP1320513205,2012RomeroIsartP147205147205, 2020GunawanP1335913359,2019OBrienP5310353103,2016HsuP3012530125,2013YinP3361433614,2020DelicP892895,2011LiP527530,2020TebbenjohannsP1360313603,2020MillenP2640126401,2020LatorreP105002105002,1989YangP461465,2017MaP2382723827,2020MillenP155168,2021PerdriatP651651,2012CirioP147206147206,2017MaP2382723827} that readily isolated the oscillators from the environment.

Optical, electrical, and magnetic levitation are the three types of suspending setups that can all work in a vacuum environment. The magnetic trap with a passive field~\cite{2019TimberlakeP224101224101, 2020VinanteP6402764027, 2020GieselerP163604163604, 2021XiongP1320513205, 2012RomeroIsartP147205147205, 2021LengP2406124061, 2019WalkerP6381963819, 2019OBrienP5310353103, 2020GunawanP1335913359, 2016JohnssonP3749537495, 2017RusconiP134419134419, 2021NavauP174436174436, 2016HsuP3012530125, 2018SlezakP6302863028} is simpler than the optical trap with lasers~\cite{2020TebbenjohannsP1360313603, 2011LiP527530, 2013YinP3361433614, 2020DelicP892895} and the electrical trap with radio-frequency modulation of a high voltage electric field~\cite{2012BrittonP489492, 2004PorrasP207901207901, 2017DelordP6381063810, 2020MartinetzP101101}. Photon recoil, damage to suspended particles caused by the laser's thermal effect, and clamping losses can all be avoided via magnetostatic field levitation~\cite{2016HsuP3012530125}. For these suspended schemes, the suspension objectives are  diverse. Glass spheres~\cite{2011LiP527530}, superconductor spheres~\cite{2020LatorreP105002105002}, superconductor rings~\cite{2021NavauP174436174436, 2012CirioP147206147206}, magnetic microspheres~\cite{2019TimberlakeP224101224101, 2017PratCampsP3400234002}, silicon particles~\cite{2020DelicP892895, 2017MaP2382723827} and diamond particles~\cite{2016HsuP3012530125, 2013YinP3361433614, 2020GunawanP1335913359, 2019OBrienP5310353103} have all been investigated on various platforms. Because of their isolation, they have been used to construct ultra-sensitive sensors~\cite{2019TimberlakeP224101224101,2017PratCampsP3400234002} as well as to couple superconducting circuits~\cite{2016JohnssonP3749537495} and solid-state spins~\cite{2021StreltsovP193602193602, 2017DelordP6381063810, 2017MaP2382723827}. Magnetic microspheres, particularly YIG (yttrium iron garnet) spheres due to their high spin density~\cite{2016BourhillP144420144420}, have received a great attention~\cite{2020GonzalezBallesteroP125404125404, 2014TabuchiP8360383603, 2020GonzalezBallesteroP9360293602, 2010SoykalP7720277202, 2015LambertP5391053910, 2016KostylevP6240262402, 2014ZhangP156401156401, 2018WangP5720257202, 2021HeiP4370643706}. There have been investigations on magnon coupling to cavity modes such as a sphere cavity~\cite{2010SoykalP7720277202}, co-axis cavity~\cite{2015LambertP5391053910}, 3D cavity~\cite{2016KostylevP6240262402}, and so on. Classical Rabi-like oscillation~\cite{2014ZhangP156401156401}, magnetically induced transparency~\cite{2014ZhangP156401156401}, bistable states~\cite{2018WangP5720257202}, and other intriguing quantum effects have been observed. Furthermore, the YIG sphere can couple to microwave photons and solid-state spins, which has been utilized to improve the coupling strength between a solid-state spin and a photon mode~\cite{2021HeiP4370643706}.  In addition, the levitated micromagnets coupled to solid-state spins have been studied~\cite{2020GieselerP163604163604}. Recent study has showed the interaction of a nitrogen-vacancy (NV) center in diamond with a levitated micromagnet through the magnetic field gradient produced by the micromagnet~\cite{2020GieselerP163604163604}. The coupling strength, however, is so weak that it can not be used for quantum information tasks.

%~\cite{2020LiP153602153602,2019TimberlakeP224101224101, 2021StreltsovP193602193602, 2020GieselerP163604163604, 2017PratCampsP3400234002}
Inspired from previous experimental and theoretical progress, we propose a useful approach to exponentially enhance the spin-mechanical coupling strength in a spin-magnetomechanical system. An NV center is situated near the hard spherical micromagnet, which levitates above a type-II superconductor. The magnetic field gradient generated by the micromagnet couples the NV center to the center-of-mass motion of the micromagnet. Many schemes have been proposed to enhance the single-quantum interaction on various platforms. Nonlinear resources~\cite{2018QinP9360193601, 2019ChenP1233912339, 2020GroszkowskiP203601203601, 2015LueP9360293602, 2016LiP1906519065} and parametric drive~\cite{2016LemondeP1133811338, 2018LerouxP9360293602} (for example, two-photon drive) have been utilized to increase light-matter interactions. The modulation of voltage in a trapped-ion system is used to achieve parametric amplification~\cite{2019GeP3050130501, 2019GeP4341743417,2021BurdP898902}. Modulating the spring constant of a cantilever~\cite{1991RugarP699702} increases the spin-phonon coupling strength exponentially in a spin-mechanical system~\cite{2020LiP153602153602}. This work suggests a classical electrical-current-driven approach for achieving exponential enhancement of spin-mechanical interactions in a suspended micromagnet platform. The driving current is located above the levitated micromagnet. The trap potential is modified by the magnetic field of the current, which modulates the oscillation frequency of the micromagnet's mechanical motion. This modulation process can provide a two-phonon drive capable of  amplifying the mechanical zero-point fluctuations, hence increasing the spin-mechanical interaction. In other words, despite merely employing a classical drive current, we obtain a nonlinear resource and, as a consequence, achieve the strong coupling regime without adding any nonlinear sources into the system. Utilizing the strongly coupled spin-mechanical system, one can prepare a high fidelity superposition state of the levitated micromagnet. In addition, the phonon-mediated spin-spin coupling can be obtained when two NV centers are coupled to the same mechanical oscillator~\cite{2009XuP2233522335, 2010ZhouP4232342323, 2013ChotorlishviliP8520185201, 2020LiP153602153602}, and the interaction can also be exponentially amplified with a driving current. With the enhanced spin-spin coupling, the two-NV protocol can also construct an unconventional 2-qubit geometric phase gate with the property of high fidelity, shorter operation time, and universality.

\section{\label{sec:II}Setup and protocol}
\subsection{\label{sec:IIA}The setup}
\begin{figure}
  \centering
  \includegraphics[width=0.48\textwidth]{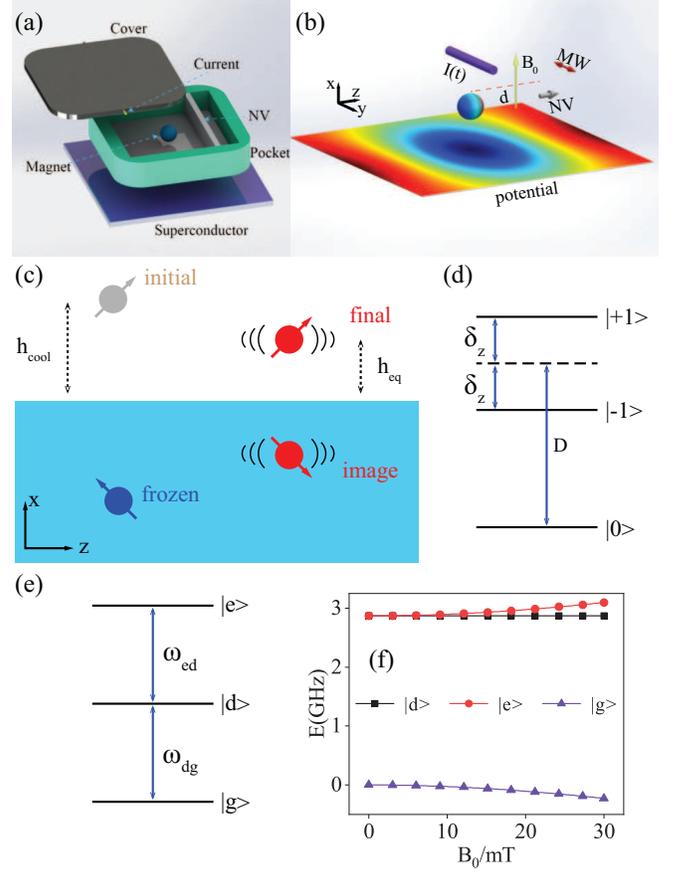}
  \caption{Setup sketch: (a) the model of our proposal. (b) shows the principle of our setup. (c) displays the frozen dipole model. (d) the energy level structure of NV centers. It will produce the Zeeman splitting when a $z$-direction magnetic field is applied. (e) the energy level structure of the mixed states. (f) the energy level splitting of the mixed states varys with the $x$-direction magnetic field.}
  \label{fig1}
\end{figure}

Fig.~\ref{fig1}(a)  presents a hybrid system that includes a micromagnet, an NV center, and a driving current. The hard spherical micromagnet with radius $a$, mass $m$, levitates on the type-II superconductor because of the flux trapping effect, the superconductor freezing or trapping the magnetic flux that penetrates it during the cooldown (see Fig.~\ref{fig1}(a))~\cite{2020GieselerP163604163604, 1998KordyukP610612, 1989YangP461465}. The microfabricated pocket provides a stable vacuum environment to isolate the micromagnet from the environment, enabling the dissipation of the system to be decreased. A cosine-function drive is provided by the current above the micromagnet, and the NV center is placed nearby the micromagnet. Fig.~\ref{fig1}(b) depicts the principle of this setup. The micromagnet trapped in the magnetostatic field, which can be calculated via  the frozen dipole model (Fig.~\ref{fig1}(c)), can be compared to a simple harmonic oscillator that couples to the NV center. The NV center transition (Fig.~\ref{fig1}(d)) is driven by a linearly polarized microwave in the $y$-direction, and the transverse static magnetic field (i.e. $x$-direction) results in a mix of the eigenstates of $\hat{\sigma}_z$. The energy level structure of the mixed states is depicted in Fig.~\ref{fig1}(e). Fig.~\ref{fig1}(f) presents the energy level splitting of the mixed states varying with the $x$-direction magnetic field.

\subsection{\label{sec:IIB}Levitation of the micromagnet}
\begin{figure}
  \centering
  \includegraphics[width=0.48\textwidth]{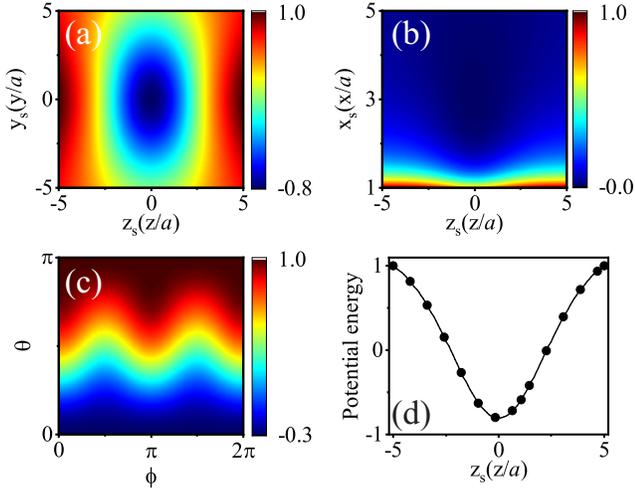}
  \caption{The potential energy of the levitated micromagnet. The dimensionless potential energy $u_s$ in $zy$-plane, $zx$-plane, $\theta \phi$-plane and $z$-direction are shown in (a), (b), (c), and (d), respectively. Here, the radius and the density of the micromagnet are $a=22.4\mu m$ and $\rho=7430 kg/m^3$, respectively. The initial position and the equilibrium position are both $3a$. The initial orientation is $\phi_{cool}=\pi/2$ and $\theta_{cool}=0$. And residual induction is $B_r = 750mT$.}
  \label{fig2}
\end{figure}

As shown in Fig.~\ref{fig1}(b), the position of the levitated micromagnet with mass $m$ and radius $a$ in the direction of gravity is represented by $x$. The acceleration of gravity is $g=9.8m/s^2$. According to the frozen dipole model~\cite{1998KordyukP610612}, the effective magnetic field $\boldsymbol{B_{eff}}$ at the position of the levitated micromagnet consists of the magnetic field $\boldsymbol{B_f}$ generated by frozen dipole and the magnetic field $\boldsymbol{B_i}$ generated by image dipole, as depicted in Fig.~\ref{fig1}(c). Then the potential energy of the levitated micromagnet is given by
\begin{equation}
  U=\bm{-\mu \cdot B_{eff}}+mgx,
\end{equation}
where the $\bm{B_{eff}}=\bm{B_f}+1/2 \bm{B_i}$ is the effective magnetic field produced by the interaction between the micromagnet and type-II superconductor~\cite{2020GieselerP163604163604}. We can derive an analytic formula for the potential energy $U$ as
\begin{equation}
  U = U_s \left(\alpha_s x_s + g_u\right),
  \label{U}
\end{equation}
where $l_s=l/a$ ($l=x,y,z$), $\alpha_s=a/\alpha_{crit}$, $U_s=mg \alpha_{crit}$, and $\alpha_{crit}=B_r^2/(16 g \rho \mu_0)$. $B_r$, $\rho$, and $\mu_0$ are  residual flux density, density of the micromagnet, and vacuum permeability. The dimensionless potential energy defined by $u_s = \alpha_s x_s + g_u$ in Eq.~(\ref{U}) is plotted in Fig.~\ref{fig2}, showing that the micromagnet can be steadily trapped in the potential trap. Fig.~\ref{fig2}(a), (b), and (c) present the dimensionless potential energy $u_s$ of the micromagnet in the $zy$-plane, $zx$-plane, and $\theta \phi$-plane, respectively. In the $zy$-direction, the potential energy exhibits strong symmetry. As depicted in Fig.~\ref{fig2}(c), the equilibrium orientation of the levitation micromagnet $\theta$ and $\phi$ is, interestingly, the same as the initial orientation $\theta_{cool}$ and $\phi_{cool}$. It means that the rotation of the micromagnet can be neglected, or that the ultimate orientation can be set as $\theta=\theta_{cool}=0$, $\phi=\phi_{cool}=\pi/2$. The potential energy distribution along the $z$-axis illustrated in Fig.~\ref{fig2}(d) can be well approximated as a harmonic potential, implying that the motion of the micromagnet is harmonic. In addition, the levitated micromagnet provides a strong magnetic field gradient for spin-mechanical coupling.

\subsection{\label{sec:IIC}Hamiltonian of the system}
The NV center is coupled to the micromagnet in this protocol through the magnetic field gradient induced by the micromagnet in the $z$-direction. In the presence of a homogeneous static magnetic field in the $x$-direction $\boldsymbol{B_s}=B_0 \boldsymbol{\hat{e}_x}$, the ground state Hamiltonian of the NV center can be written as $\hat{H}_{NV}=D\hat{S}_z^2+\gamma_e\bm{\hat{S} \cdot B_s}$, where $\gamma_{e}=g_e\mu_B$ is the electron gyromagnetic factor and $D/2\pi=2.87$~GHz is the zero-field splitting between sublevel $\vert m_s=0 \rangle$ and $\vert m_s=\pm 1 \rangle$ (see Fig.~\ref{fig1}(d)). $g_e\simeq 2$ and $\mu_B=14$MHz/mT are the Landa\'e factor of electron and Bohr magneton, respectively. $\boldsymbol{\hat{S}}$ is electron spin operator including the components $\hat{S}_x$, $\hat{S}_y$, and $\hat{S}_z$. The microwave (MW) drive $B_y(t)=B_y^0 \cos \omega_p t$ polarized in the $y$-direction is applied to drive the transition between the sublevels, where $B_y^0$ and $\omega_p$ are the amplitude and the microwave frequency of the drive respectively. Then the Hamiltonian is given by
\begin{equation}
  \hat{H}_{NV}=D\hat{S}_z^2+\delta \hat{S}_x + \Omega_p {\rm cos} \left(\omega_p t \right) \hat{S}_y,
  \label{eq4}
\end{equation}
where $\Omega_p=\gamma_e B_{y}^{0}$ is the Rabi frequency and $\delta=\gamma_e B_0$.

The motion of a levitated micromagnet can be regarded as three independent harmonic motions in three directions around the equilibrium position. Only the harmonic motion of the $z$-direction, which is the same as the spin direction of the NV center, is considered here. Its Hamiltonian is
\begin{equation}
  \hat{H}_{ma}=\frac{\hat{p}_z^2}{2m}+\frac{1}{2}k_{ma} \hat{z}^2
\end{equation}
with
\begin{equation}
  k_{ma}=\frac{3 \mu_0 \mu_m^2}{4\pi (h_{eq}+h_{cool})^5},
\end{equation}
where $\hat{p}_z$ and $\hat{z}$ are the momentum and position operators, respectively, and $\mu_m$ is the magnetic moment of the micromagnet. $h_{eq}$ and $h_{cool}$ represent the initial position and equilibrium position of the micromagnet respectively. The trapping frequency of the levitated micromagnet is defined by $\omega_{ma}=\sqrt{k_{ma}/m}$, which is related to the cooling down conditions.  With $\hat{p}_z=-i\sqrt{m\omega_{ma}/2} (\hat{a}-\hat{a}^{\dagger})$, $\hat{z}=z_0 (\hat{a}+\hat{a}^{\dagger})$, and zero-point fluctuation $z_0=\sqrt{1/(2m \omega_{ma})}$, we can simplify the Hamiltonian of the micromagnet. The result is
\begin{equation}
  \hat{H}_{ma}=\omega_{ma} \hat{a}^{\dagger} \hat{a},
  \label{eq6}
\end{equation}

The interaction between the NV center and the micromagnet will be the subject of our next discussion. The micromagnet can be described as a magnetic dipole in classical electrodynamics, with
\begin{equation}
  \boldsymbol{B}=\frac{\mu_0}{4\pi} \left[\frac{3 \boldsymbol{r}(\boldsymbol{\mu} \cdot \boldsymbol{r})}{r^{5}}-\frac{\boldsymbol{\mu}}{r^{3}}\right]
\end{equation}
describing the magnetic field surrounding it. Only the magnetic field in the $z$-direction is concerned here, which is given by $\bm{B}= 2 \mu_0 \mu_m \bm{\hat{e}_z}/ [4\pi  (d-z )^3 ]$. The magnetic field is represented by
\begin{equation}
  \bm{B}\simeq \frac{2\mu_0 \mu_m \bm{\hat{e}_z}}{4 \pi d^3}+ \frac{3\mu_0 \mu_m \hat{z} \bm{\hat{e}_z}}{4 \pi d^4}+O(\hat{z}^2)
\end{equation}
around the equilibrium position. After removing the high order and constant components, and quantizing the motion (more details in Appendix~\ref{sec:appendixB}), the interaction Hamiltonian is expressed as
\begin{equation}
  \hat{H}_{int}=\lambda (\hat{a} + \hat{a}^{\dagger})  \hat{S}_z,
  \label{eq7}
\end{equation}
where $\lambda=\gamma_e B_r a^3 z_0 / d^4$ is the coupling strength and $d$ is the distance between the NV center and the micromagnet.

Finally, $I_{cu}=I_0 {\rm cos} 2\omega_{cu} t$ is the drive current placed above the micromagnet, where $\omega_{cu}$ is the driving current frequency and $I_0$ is the amplitude of the driven electrical current. The Hamiltonian of the driving current is given by $\hat{H}_{cu}=1/2 k_{cu}^2 \hat{z}^2 {\rm cos} 2 \omega_{cu} t$. After quantization (Appendix~\ref{sec:appendixC}), we can have
\begin{equation}
  \hat{H}_{cu}=-g_{cu} (\hat{a} + \hat{a}^{\dagger})^2 {\rm cos} 2\omega_{cu}t ,
  \label{eq8}
\end{equation}
where $g_{cu}=k_{cu} z_0^2 /2$ defines the coupling strength between the driving current and the micromagnet. The nonlinear term or the parametric amplification is obtained by the linear drive. The spin-mechanic coupling strength can be exponentially enhanced with such a nonlinear term, as demonstrated below.
\section{\label{sec:III}enhancing the coupling strength}
\subsection{\label{sec:IIIA}One NV}
Based on the foregoing analysis, the total Hamiltonian of the hybrid system is
\begin{equation}
  \hat{H}_{TO} = \hat{H}_{NV} + \hat{H}_{ma} + \hat{H}_{int} + \hat{H}_{cu}.
\end{equation}
The first term is the Hamiltonian of NV centers. The second corresponds to the free Hamiltonian of the micromagnet, the third describes the interaction between the NV center and micromagnet, and the last is the drive-current Hamiltonian. In the absence of the microwave drive, the Hamiltonian $\hat{H}_{NV}$ is given by
\begin{equation}
  \hat{H}_{NV} = D \hat{S}_z^2 + \delta \hat{S}_x.
  \label{eqD1}
\end{equation}
The eigenstates of Eq.~(\ref{eqD1}) are mixed states $\vert e \rangle={\rm sin} \theta \vert 0 \rangle + {\rm cos} \theta \vert b \rangle$, $\vert g \rangle={\rm cos} \theta \vert 0 \rangle - {\rm sin} \theta \vert b \rangle$, and $\vert d \rangle= \left(\vert +1\rangle - \vert -1\rangle \right)/\sqrt{2}$, where $\vert b \rangle= \left(\vert -1\rangle + \vert +1\rangle \right)/\sqrt{2}$ and ${\rm tan} 2\theta = 2\delta / D$, corresponding to the eigenenergy $\omega_{e/g}=D[1\pm \sqrt{1+(2\delta/D)^2}]/2$, and $\omega_{d}=D$.

\begin{figure}
  \centering
  \includegraphics[width=0.48\textwidth]{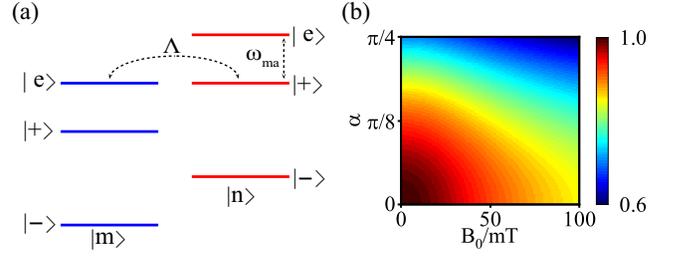}
  \caption{Spin-phonon coupling. (a) shows couplings between the spin and phonons. The intensity of the spin-phonon coupling as a function of $\alpha$ and the transverse static magnetic field $B_0$ is shown in (b).}
  \label{fig3}
\end{figure}

We assume that the microwave is solely used to drive the transition between the mixed states $\vert g \rangle$ and $\vert d \rangle$, i.e. $\omega_p \simeq \omega_d - \omega_g = \omega_{dg}$. Transforming to the frame at the microwave frequency and using the rotating-wave approximation, the Hamiltonian in the basis $\vert e, d, g \rangle$ of the NV center can be reduced as
\begin{equation}
  \hat{H}_{N V}=\left(\begin{array}{ccc}
    -\frac{\Delta}{2} & 0 & 0 \\
    0 & \frac{\Delta}{2} & 0 \\
    0 & 0 & \omega_{e}^{\prime \prime}
    \end{array}\right)+\frac{1}{2 i}\left(\begin{array}{ccc}
    0 & -\Omega_{p}^{\prime} & 0 \\
    \Omega_{p}^{\prime} & 0 & 0 \\
    0 & 0 & 0
    \end{array}\right),
  \label{eq12}
\end{equation}
where $\Delta=\omega_p-\omega_{dg}$, $\omega_e^{\prime \prime}=\omega_e- (\omega_d + \omega_g + \omega_p)/2$, and $\Omega_p^{\prime}=\Omega_p cos \theta$. We consider a new basis for further diagonalization, consisting of the eigenstates of Eq.~(\ref{eq12}), which are $\vert e \rangle$, $\vert +\rangle = i \sin \alpha \vert g \rangle + {\cos} \alpha \vert d \rangle$, and $\vert -\rangle = -i {\cos} \alpha \vert g \rangle + {\sin} \alpha \vert d \rangle$, with eigenenergies $\omega_e^{\prime \prime}$ and $\omega_{\pm}=\pm \sqrt{\Delta ^ 2+\Omega_p^{\prime 2}} / 2$, where ${\tan} 2\alpha=\Omega_p^{\prime}/\Delta$. Using the new eigen basis ${\vert e, \pm \rangle}$ and the unitary transformation $U={\rm exp} (-i \hat{H}_u t)$ with $\hat{H}_u=\omega_{cu} (\hat{\sigma}_z / 2 + \hat{a} ^ {\dagger} \hat{a} )$, the Hamiltonian of the hybrid system is represented as
\begin{equation}
  \begin{aligned}
    \hat{H}_{TO}=\frac{\delta_{0}}{2} \hat{\sigma}_{z}+\delta_{m} \hat{a}^{\dagger} \hat{a}+\Lambda\left(\hat{a} \hat{\sigma}^{+}+\hat{a}^{\dagger} \hat{\sigma}^{-}\right) \\
    -\frac{g_{c u}}{2}\left(\hat{a}^{2}+\hat{a}^{\dagger^{2}}\right),
  \end{aligned}
  \label{eq14}
\end{equation}
where $\hat{\sigma}_z \equiv \vert e \rangle \langle e \vert - \vert + \rangle \langle + \vert $, $\hat{\sigma}_x=\hat{\sigma}_{+} + \hat{\sigma}_{-}$, $\hat{\sigma}_{+} \equiv \vert e \rangle \langle + \vert $, $\hat{\sigma}_{-} \equiv \vert + \rangle \langle e \vert$, $\delta_0=\omega_0-\omega_{cu}$, $\omega_{0}=\omega_{e}^{\prime \prime}-\omega_+$, and $\delta_m=\omega_{ma}-\omega_{cu}$. Here only the states  $\vert m, + \rangle$ and $\vert n, e \rangle$ are resonant with the condition $\left\lvert m-n \right\rvert = 1$, $m$ and $n$ being the phonon numbers (see Fig.~\ref{fig3}(a)). Under the aforesaid resonant condition, the spin-phonon coupling strength is given by $\Lambda=\lambda {\rm cos} \theta {\rm cos} \alpha$, which is related to the transverse magnetic field $B_0$ and $\alpha$, dependent on the microwave frequency $\Omega_p^{\prime}$ and $\Delta$, as shown in Fig.~\ref{fig3}(b). The coupling strength increases as $B_0$ and $\alpha$ decrease, showing that we should choose an appropriate value to make the system work well.

\begin{figure}
  \centering
  \includegraphics[width=0.48\textwidth]{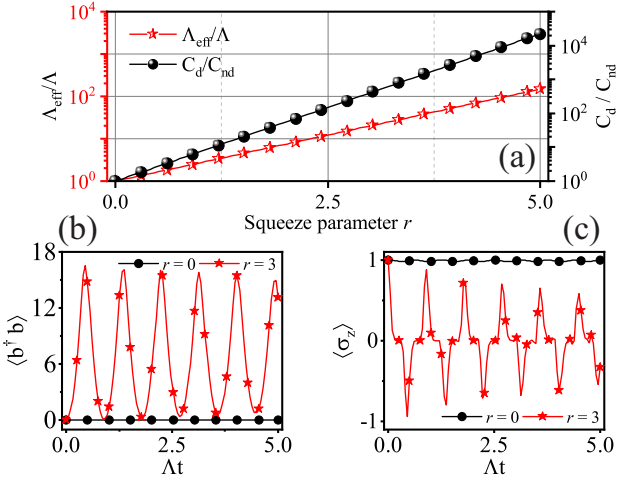}
  \caption{Spin-phonon and phonon-mediated spin-spin coupling. (a) shows $\Lambda_{eff} / \Lambda$ and $C_{d} / C_{nd}$ as a function of the squeezing parameter $r$. The spin-phonon coupling strength appears to be enhanced exponentially. The dynamic processes with ($r=3$) and without ($r=0$) the drive current are shown in (b) and (c). Here the initial state is $\vert 0 \rangle \vert \uparrow \rangle$, where the $\vert \uparrow \rangle$ represents the excited state $\vert e \rangle$. The coefficients are $\delta_m=10\Lambda$, $\delta_0=0$, and $r=0,3$. The dephasing of NV centers and the dissipation of the micromagnet are $\gamma_{NV}=0.01\Lambda$ and $\kappa_{ma}=0.01\Lambda$, respectively.}
  \label{fig4}
\end{figure}

Using the Bogoliubov transformation~\cite{2016LemondeP1133811338, 2019BurdP11631165, 2021BurdP898902} $\hat{b}=\hat{a}^{\dagger} {\rm cosh} r - \hat{a} {\rm sinh} r$, with ${\rm tanh} 2r=g_{cu}/\delta_m$, the total Hamiltonian can be expressed in a simple form
\begin{align}
  &\hat{H}_{TO} = \hat{H}_{RO} + \hat{H}_{Sq} \label{HTO}, \\
  &\hat{H}_{RO}=\frac{\delta_{0}}{2} \hat{\sigma}_{z}+\Delta_{m} \hat{b}^{\dagger} \hat{b}+\Lambda_{eff}\left(\hat{b} + \hat{b}^{\dagger}\right) \hat{\sigma}_{x} \label{HRO}, \\
  &\hat{H}_{S q}=\frac{\Lambda e^{-r}}{2}\left(\hat{b} - \hat{b}^{\dagger}\right)\left(\hat{\sigma}^{-}-\hat{\sigma}^{+}\right) \label{HSq},
\end{align}
where $\Lambda_{eff}=\Lambda e^r / 2$ and $\Delta_m = \delta_m/{\rm cosh} 2r$. $\hat{b}$ ($\hat{b}^{\dagger}$) is the annihilation (creation) operator of Bogoliubov modes. Because of the driving current, the spin-phonon coupling strength can be enhanced exponentially. The spin-phonon coupling strength is orders of magnitude larger than the original one, as seen in Fig.~\ref{fig4}(a). Because the item $e^{-r}$ decreases to zero as the squeezing parameter $r$ increases, the term $\hat{H}_{Sq}$ can be ignored.

To quantify the spin-mechanic coupling strength the cooperativity $C_{nd}=\Lambda^{2}/ (\kappa_{ma}\gamma_{NV})$, a dimensionless parameter, is introduced, where $\kappa_{ma}$ and $\gamma_{NV}$ are the mechanical dissipation and the spin dephasing, respectively. Inevitably, as the coupling strength is amplified, so is the mechanical noise. To alleviate the negative consequences of amplified mechanical noise, the dissipative squeezed scheme proposed in the literature~\cite{2020LiP153602153602, 2015PirkkalainenP243601243601, 2015WollmanP952955} can be used. Through the dissipative squeezed method, the $b$-mode is always in the ground state in the squeezed picture. In this case, the Lindblad master equation of the system can be expressed as
\begin{equation}
  \dot{\hat{\rho}}=-i [\hat{H}_{RO}, \hat{\rho}]+\kappa_{ma}^{au}D (\hat{b}) \hat{\rho}+\gamma_{NV} D (\hat{\sigma}_z) \hat{\rho},
  \label{Hnoise}
\end{equation}
where $D(\hat{O})\hat{\rho}=\hat{O}\hat{\rho}\hat{O}^{\dagger}-\{\hat{O}^{\dagger}\hat{O}, \hat{\rho}\}/2$ is the Lindblad operator, and $\kappa_{ma}^{au}$ is the effective mechanical dissipation resulting from the interaction between the mechanical mode and the auxiliary bath. And then effective cooperativity with the driving current is given by $C_{d}  = \Lambda_{eff}^{2}/ (\kappa_{ma} ^ {au} \gamma_{NV})$. As a result, we can get $ C_{d} / C_{nd} \sim e^{2r}$, which is magnified exponentially, as shown in Fig.~\ref{fig4}(a). Using the master equation (\ref{Hnoise}), we numerically evaluate the dynamic processes with and without the driving current. In the absence of the driving current, the coupling strength between spins and phonons is extremely weak, resulting in no Rabi oscillation; in the presence of the driving current, the spin-phonon coupling strength is greatly  enhanced, resulting in Rabi oscillations, as illustrated in Fig.~\ref{fig4}(b) and (c). To put it in other words, the driving current can be employed to enhance the spin-phonon coupling.
\subsection{\label{sec:IIIB}Geometric phase}
\begin{figure}
  \centering
  \includegraphics[width=0.48\textwidth]{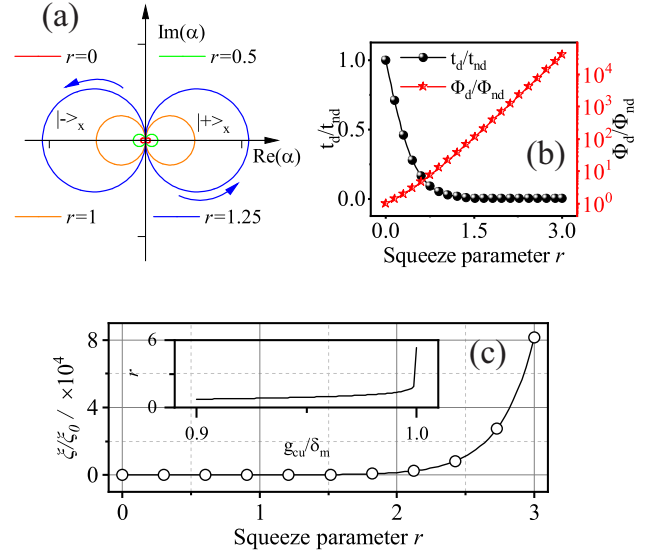}
  \caption{In (a), the phonon trajectory in phase space is depicted, which encloses a larger area as the squeeze parameter increases. (b) geometric phase (in one period) and time  (required to acquire a given geometric phase) are depicted as a function of the squeezing parameter $r$. (c) shows that the phonon-mediated spin-spin coupling strength varies with the squeezing parameter $r$. The squeezing parameter $r$ is plotted as a function of the ratio $g_{cu}/\delta_m$ in the inset.}
  \label{fig5}
\end{figure}

Now we focus on the process of enhancement in phase space. Considering Eq.~(\ref{HRO}), for the sake of simplicity, we set $\delta_0=0$ and move into the Bogoliubov-mode interaction representation. The time evolution operator of the system $U_{RO}(t) = D[\alpha (t)] {\rm exp} [i \Phi (t)\hat{\sigma}_x^2]$ is obtained via Magnus expansion~\cite{2006ZhuP485491, 2008RoosP1300213002, 2018ArnalP3502435024, 2009BlanesP151238}, where  $D[\alpha (t)]={\rm exp}{[\alpha (t) \hat{b}^{\dagger} - \alpha^{*} (t) \hat{b}] \hat{\sigma}_x}$ is the displacement operator and $\alpha (t)= \Lambda_{eff}/\Delta_m (1 - e^{i \Delta_m t})$ is the coherent displacement of the phonon in phase space.  The spin and phonon are decoupled at time $t=2\pi N/\Delta_m$ with $N=1,2,3,\cdots$, as shown by the time evolution operator $U_{RO}(t)$, and the phonon returns to its initial state. In Fig.~\ref{fig5}(a), the phonon-mode trajectory is shown in phase space. Due to the driving current, the phase space trajectory is magnified and covers a broader area. In addition, the phonon migration direction in phase space is correlated to the spin state, as indicated in equation $D[\alpha (t)]$. Under the original representation, i.e., the interaction representation of phonons, the phase space displacement of phonons is written as $\alpha_I (t)=\Lambda / (2\Delta_m) [({\rm cos} \Delta_m t - 1 )e^{2r} - i {\rm sin} \Delta_m t]$ ~\cite{2019GeP3050130501, 2021BurdP898902}.

The geometric phase $\Phi$ is determined only by the enclosed area swept away by phonon trajectories in phase space, as given by
\begin{equation}
  \Phi=Im \left[\int_{0}^{t} \alpha^{*} (t^{\prime})\,d\alpha (t^{\prime}) \right].
\end{equation}
The geometric phase with the driving current at time $t=2\pi /\Delta_m$ (phonons orbit once in phase space) is given by $\Phi_d=2\pi (\Lambda_{eff}/\Delta_m)^2$. In Fig.~\ref{fig5}(b), $\Phi_d /\Phi_{nd} \varpropto  (e^r {\rm cosh} 2r) ^ 2$ is shown as a function of the squeezing parameter $r$, where $\Phi_{nd}=2\pi (\Lambda/\delta_m)^2$ is the geometric phase sans drive. The geometric phase is roughly exponentially increased. We currently consider acquiring a certain geometric phase $\Phi_0$. After that, we can get $ t_d /t_{nd} \varpropto 1/ (e^r {\rm cosh} 2r)$, where $t_{d}$  and $t_{nd}$ are the time required to acquire $\Phi_0$ with and without a drive, respectively. As seen in Fig.~\ref{fig5}(b), increasing the squeezing parameter $r$ reduces the time required to acquire a given phase $\Phi_0$.

\subsection{\label{sec:IIIC}Two NV}
We now discuss the interaction of two NV centers with a micromagnet. Two NVs are symmetrically arranged on either side of the micromagnet along the magnetic field direction, coupling to the micromagnet center of mass motion via a strong magnetic field gradient. In the squeezing frame (i.e. with the Bogoliubov transformation~\cite{2016LemondeP1133811338, 2019BurdP11631165, 2021BurdP898902} $\hat{b}=\hat{a}^{\dagger} {\rm cosh} r - \hat{a} {\rm sinh} r$ and ${\rm tanh} 2r=g_{cu}/\delta_m$), the Hamiltonian of the hybrid system consisting of the NVs and micromagnet is given by
\begin{equation}
  \begin{aligned}
    \hat{H}_{RT}=\frac{\delta_{0}}{2}\left(\hat{\sigma}_{z}^{1}+\hat{\sigma}_{z}^{2}\right)+\Delta_{m} \hat{b}^{\dagger} \hat{b} \\
    +\Lambda_{e f f}\left(\hat{b} + \hat{b}^{\dagger}\right)\left(\hat{\sigma}_{x}^{1}-\hat{\sigma}_{x}^{2}\right).
  \end{aligned}
  \label{eq22}
\end{equation}
(The complete derivation is given in Appendix~\ref{sec:appendixD}.)  With $\delta_0=0$, the Hamiltonian can be reduced using the Schrieffer-Wolff transformation~\cite{2004WilsonRaeP7550775507, 2013AlbrechtP8301483014} $\hat{H}_{RT}^{eff}=e^S\hat{H}_{RT}e^{-S}$, where $S=\eta (\hat{b}^{\dagger}-\hat{b})  (\hat{\sigma}_x^1 - \hat{\sigma}_x^2)$ and $\eta=\Lambda_{eff}/\Delta_m$. It is worth noting that the parameter $\eta$ is usually much smaller than one, indicating that it satisfies the Lamb-Dicke condition $\eta \ll 1$, which is similar to that for trapped ions~\cite{2020WangP230501230501}. The effective Hamiltonian is given by
\begin{equation}
  \hat{H}_{RT}^{eff} = \Delta_m \hat{b}^{\dagger} \hat{b} \\
  - \xi \left(\hat{b} + \hat{b} ^ {\dagger}\right) \left(\hat{\sigma}_x^1-\hat{\sigma}_x^2\right)^2,
  \label{eq23}
\end{equation}
where $\xi = \Lambda_{eff}^2/\Delta_m$. Retaining only the terms containing $\xi$, we obtain the Ising interaction Hamiltonian
\begin{equation}
  \hat{H}_{Ising} = \xi \left(\hat{\sigma}_x^1-\hat{\sigma}_x^2\right)^2,
  \label{eq24}
\end{equation}
corresponding to the one-axis twisting interaction~\cite{1993KitagawaP51385143}. In this scenario, the effective spin-spin interaction of the two NVs is obtained, and the phonon is only virtually excited. Fig.~\ref{fig5}(c) shows the coupling strength between two NVs and the insert depicts the squeezing parameter $r$ as a function of $g_{cu}/\delta_m$. The ratio of the amplified spin-spin coupling ($\xi = \Lambda_{eff}^2/\Delta_m$) to the bare coupling ($\xi_0 = \Lambda^{2}/\delta_m$), given by $\xi / \xi_0 \varpropto (e^{2r} {\rm cosh} 2r)^2$, exponentially increases. The phonon-mediated spin-spin interaction can be enhanced up to several orders of magnitude stronger than the bare coupling, as the squeezing parameter $r$ increases. It is independent of the specific frame of the phonon since the phonon mode has been adiabatically eliminated. The spin-spin interaction is at the heart of several quantum technologies, such as qubit gates, which are vital for quantum computer implementation. In part~\ref{sec:IVB}, we will consider a 2-qubit gate with excellent fidelity and faster gate speed.
\section{\label{sec:IV}application}
\subsection{\label{sec:IVA}Preparing Schr\"odinger cat states}
\begin{figure}
  \centering
  \includegraphics[width=0.48\textwidth]{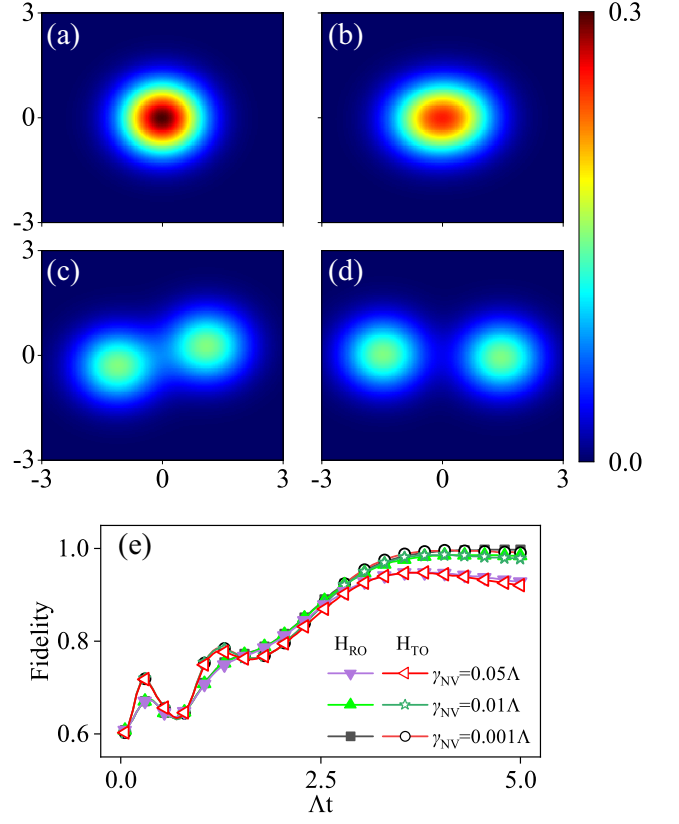}
  \caption{Schr\"odinger cat states. With $\kappa_{ma}=\gamma_{NV}=0.001\Lambda$, the Winger functions of the phonon mode are displayed in (a), (b), (c), and (d), corresponding to the situations $\Lambda t=0$, $\Lambda t=1.5$, $\Lambda t=3$, and $\Lambda t=4.5$, respectively. (e) The fidelity of the cat state is evaluated with different dephasing rates of the NV center. Furthermore, the dynamic process resulting from $\hat{H}_{RO}$ is identical to that resulting from $\hat{H}_{TO}$. Here, $r(t)=r_{max} \tanh(\Lambda t / 2)$ where $r_{max}=1.25$.}
  \label{fig6}
\end{figure}
The single NV hybrid system can be utilized to prepare a Schr\"odinger cat state~\cite{2014AsjadP4550245502,2017DasP3383733837}, which is a linear superposition of two coherent states. According to the analysis of part~\ref{sec:IIIA}, the coupling strength of the NV center and micromagnet has been greatly enhanced, which is critical for preparing a cat state with the spin-mechanical  interaction. We assign $\delta_0=0$ for the Hamiltonian (\ref{eq14}). The Hamiltonian can be diagonalized with the Bogoliubov transformation~\cite{2016LemondeP1133811338, 2019BurdP11631165, 2021BurdP898902} $\hat{b}=\hat{a}^{\dagger} {\rm cosh} r (t) - \hat{a} {\rm sinh} r (t)$ with ${\rm tanh} 2r(t)=g_{cu}(t)/\delta_m$, which reads
\begin{align}
  &\hat{H}_{TO}=\hat{H}_{RO} + \hat{H}_{S q} + \hat{H}_{E r r} , \label{HTOt} \\
  &\hat{H}_{RO}=\Delta_{m} \left(t\right) \hat{b}^{\dagger} \hat{b} + \Lambda_{e f f} \left(t\right) \left(\hat{b} + \hat{b}^{\dagger}\right) \hat{\sigma}_{x} , \label{HROt} \\
  &\hat{H}_{S q}=\frac{\Lambda e^{-r(t)}}{2}\left(\hat{b} - \hat{b}^{\dagger}\right)\left(\hat{\sigma}^{-}-\hat{\sigma}^{+}\right) , \label{HSqt} \\
  &\hat{H}_{E r r}=-i \frac{\dot{r}(t)}{2}\left(\hat{b}^{\dagger^{2}} - \hat{b}^{2}\right),
  \label{HErrt}
\end{align}
where $\Lambda_{eff} \left(t\right)=\Lambda e^{r (t)} / 2$ and $\Delta_m \left(t\right) = \delta_m/{\rm cosh} 2r (t)$. $\hat{b}$ ($\hat{b}^{\dagger}$) corresponds to the annihilation (creation) operator of the Bogoliubov mode. The Hamiltonian Eq.~(\ref{HROt}) is the time-dependent Rabi model, and the undesirable corrections are $\hat{H}_{Sq}$ and $\hat{H}_{Err}$. The Hamiltonian $\hat{H}_{Sq}$ can be ignored since it contains $e^{ - r (t)}$, as previously stated. We assume that the pump varies slowly over time to maintain adiabaticity during the dynamical process, such that the correction item $\hat{H}_{Err}$ can be ignored because $\dot{r (t)} \approx 0$. Utilizing Magnus expansion~\cite{2006ZhuP485491, 2008RoosP1300213002, 2018ArnalP3502435024, 2009BlanesP151238}, then, the time evolution operator can be written as $U_{RO}(t) = D[\alpha (t)] {\rm exp}[-i \chi (t, 0) \hat{b}^{\dagger} \hat{b}]$, where $D[\alpha (t)]={\rm exp}{[\alpha (t) \hat{b}^{\dagger} - \alpha^{*}(t) \hat{b} ] \hat{\sigma}_x }$ is the displacement operator and $\alpha (t)= -i \Lambda / 2 \int_{0}^{t} {\rm exp} [r (t^{\prime}) - i \chi (t, t^{\prime})] \,dt^{\prime} $ is the coherent displacement of phonons in phase space, with $\chi (t, t^{\prime}) =\int_{t^{\prime}}^{t} \Delta_m (t^{\prime \prime}) \,d t^{\prime \prime} $. The spin-mechanical system is prepared in the initial state $\vert \Psi_0 \rangle =\vert 0 \rangle \vert \downarrow \rangle$, with $\vert \downarrow \rangle$ representing the ground state, and the time evolution operator is then applied to the initial state. Finally, we can obtain an entangled cat state,
\begin{equation}
  \Psi_{final} = \frac{\vert \alpha \left( t \right)\rangle \vert + \rangle _x - \vert -\alpha \left( t \right) \rangle \vert - \rangle _x}{\sqrt{2}},
  \label{28}
\end{equation}
where the states $\vert \pm \alpha (t) \rangle$ are the phonon mode coherent states, and $\vert \pm \rangle _x = (\vert \uparrow \rangle \pm \vert \downarrow \rangle)/\sqrt{2}$ are the eigenstates of the operator $\hat{\sigma}_x$, with $\vert \uparrow \rangle$ being the excited state $\vert e \rangle$. From $t=0$ to $t=t_f$, the ideal Rabi Hamiltonian Eq.~(\ref{HROt}) and the total Hamiltonian Eq.~(\ref{HTOt}) are used to carry out the dynamic simulations of the aforementioned process, respectively. If we assume that the initial state is $\vert 0 \rangle \vert \downarrow \rangle$, the spin dephasing is $\gamma_{NV}$, and the phonon dissipation is $\kappa_{ma}$, the dynamic evolution follows the Lindblad master equation
\begin{equation}
  \dot{\hat{\rho}}=-i[\hat{H}_{RO} / \hat{H}_{TO}, \hat{\rho}] + \gamma_{NV} D\left(\hat{\sigma}_{z}\right) \hat{\rho} + \kappa_{ma} D(\hat{b}) \hat{\rho},
\end{equation}
where $D(\hat{O})\hat{\rho}=\hat{O}\hat{\rho}\hat{O}^{\dagger}-\{\hat{O}^{\dagger}\hat{O}, \hat{\rho}\}/2$ is the Lindblad operator. Figs.~\ref{fig6}(a), (b), (c), and (d) depict the evolution of the phonon-mode Winger function over time using the Hamiltonian $\hat{H}_{RO}$. At the initial time $t=0$, the squeezed parameter $r(0)=0$, indicating that the current drive is zero, and the system is prepared in the initial state $\vert 0 \rangle \vert \downarrow \rangle$~\cite{2018LerouxP9360293602,2020LiP153602153602}. The current drive is loaded adiabatically over time and then, the transformed-$b$-mode evolves into a well-separated Schr\"odinger cat state in phase space. In addition, the fidelity of the cat state is depicted in Fig.~\ref{fig6}(e) (with the Hamiltonian $\hat{H}_{RO}$), achieving $99.7\%$ when $\gamma_{NV}=0.001 \Lambda$, $98.3\%$ when $\gamma_{NV}=0.01 \Lambda$, and $92.8\%$ when $\gamma_{NV}=0.05 \Lambda$. It is worth noticing that the evolution predicted by $\hat{H}_{RO}$ (solid lines with open symbols) matches that predicted by $\hat{H}_{TO}$ (the solid line with close symbols). It suggests that the unwanted corrections produced by $\hat{H}_{Sq}$ and $\hat{H}_{Err}$ can be ignored.

\subsection{\label{sec:IVB}Two-qubit gate}
Quantum logic gates~\cite{2019ChenP2231922319, 2018LeungP2050120501, 2008RoosP1300213002, 2005LeeP371383, 2003LeibfriedP412415} are the core of quantum computation. Geometric quantum computing refers to the quantum computation associated with the pure geometric phase~\cite{2006JoshiP390395, 2000FalciP355358}. Based on the different methods of obtaining geometric phase, the geometric phase gate can be divided into two categories: (i) the conventional geometric phase gate, which acquires the pure geometric phase with adiabatic evolution of qubits; and (ii) the unconventional geometric phase gate, which acquires the pure geometric phase with the evolution of the bose mode along a close trajectory in the phase space~\cite{2006JoshiP390395, 2004ZhengP5232052320}. Conventional geometric phase gates have been studied with many platforms~\cite{2000FalciP355358, 2000JonesP869871, 2001DuanP16951697}. The two-NV proposal, as discussed in part~\ref{sec:IIIC}, can be utilized to build an unconventional geometric phase 2-qubit gate with high fidelity and faster gate speed. The hybrid system containing two NVs is described by the Hamiltonian (\ref{eq22}). As previously discussed, we set $\delta_0=0$, and $\hat{b}$ ($\hat{b}^{\dagger}$) corresponds to the Bogoliubov mode annihilation (creation) operator. Moving in the Bogoliubov-mode interaction frame, we can get
\begin{equation}
  \hat{H}_{RT}^I = \Lambda_{eff} \left(\hat{b} e^{-i \Delta_m t} + \hat{b} ^ {\dagger}e^{i \Delta_m t}\right) \left(\hat{\sigma}_x^1 - \hat{\sigma}_x^2\right).
\end{equation}
Then, utilizing the Magnus expansion~\cite{2006ZhuP485491, 2008RoosP1300213002, 2018ArnalP3502435024, 2009BlanesP151238}, the time evolution operator is given by
\begin{equation}
  U_{RT} (t) = D[\alpha (t)] E_{ij}[\beta (t)],
\end{equation}
where $D[\alpha (t)]={\rm exp}{[\alpha (t) \hat{b}^{\dagger} - \alpha^{*} (t) \hat{b}] \hat{\sigma}_x }$ denotes the displacement operator, and $\alpha (t)= \Lambda_{eff}/\Delta_m (1-e^{i \Delta_m t})$ is the coherent displacement of phonons in phase space. The second item describing spin-spin interaction is given by
\begin{equation}
  E_{i j}[\beta(t)]={\rm exp} \left(\sum_{i, j}^{2} \beta \eta_{i j} \hat{\sigma}_{x}^{i} \hat{\sigma}_{x}^{j}\right),
\end{equation}
where
\begin{equation}
  \eta_{i j}=\left\{\begin{array}{c}
    1, i=j \\
    -1, i \neq j
    \end{array}\right. .
\end{equation}
The phonon mode returning to its initial state, a gate operation is completed. As a result, the gate time is determined by $\tau=2\pi/\Delta_m$, at which point the time evolution operator can be represented as
\begin{equation}
  U_{RT} (\tau) = {\exp} \left(-i 2 \pi \frac{\Lambda_{eff}^2}{\Delta_m^2} \sum_{i,j}^{2} \eta _{ij} \hat{\sigma}_x^i \hat{\sigma}_x^j\right).
\end{equation}
Adjusting the ratio between $\Lambda_{eff}$ and $\Delta_m$, qubit gates corresponding to different phases can be constructed, such as the $\pi / 2$-2-qubit gate described by $U_{RT} (\tau) = {\rm exp} (-i \frac{\pi}{8} \sum_{i,j}^{2} \eta _{ij} \hat{\sigma}_x^i \hat{\sigma}_x^j)$. Supposing that the initial state is the eigenstate of $\hat{\sigma}_x$, then, at time $\tau$, the final state is
\begin{equation}
  \begin{array}{c}
    \vert + \rangle_x \vert + \rangle_x \rightarrow \vert + \rangle_x \vert + \rangle_x ,\\
    \vert - \rangle_x \vert - \rangle_x \rightarrow \vert - \rangle_x \vert - \rangle_x ,\\
    \vert + \rangle_x \vert - \rangle_x \rightarrow e^{-i\pi / 2} \vert + \rangle_x \vert - \rangle_x , \\
    \vert - \rangle_x \vert + \rangle_x \rightarrow e^{-i\pi / 2} \vert - \rangle_x \vert + \rangle_x .
  \end{array}
  \label{34}
\end{equation}
The 2-qubit gate only adds phase to $\vert + \rangle _x \vert - \rangle _x $ and $\vert - \rangle _x \vert + \rangle _x $, not $\vert + \rangle _x \vert + \rangle _x $  and $\vert - \rangle _x \vert - \rangle _x $, because the time evolution operator at time $\tau$ is $\hat{I}$ when the initial state is the latter. Furthermore, the 2-qubit gate is universal, as demonstrated in the literature~\cite{1995DeutschP669677}.
\begin{figure}
  \centering
  \includegraphics[width=0.48\textwidth]{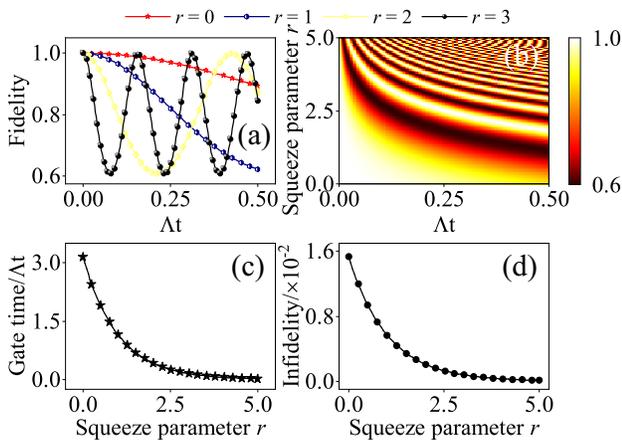}
  \caption{2-qubit gate. The dynamic process of the $2$-qubit gate is presented in (a), (b) with the varying squeezing parameter $r$. When the phonon mode evolves back to the original state, a gate operation is accomplished. In (c) and (d), gate time and the infidelity of the $2$-qubit gate are shown as a function of the squeezing parameter $r$. The larger the squeezing parameter $r$, the shorter the gate-time and the smaller the gate infidelity. Here, the parameters are $\kappa_{ma}=\gamma_{NV}^1=\gamma_{NV}^2=0.01\Lambda$, $\delta_0=0$, $\Delta_m=4\Lambda_{eff}$ and the initial state is $\vert + \rangle_x \vert - \rangle_x$.}
  \label{fig7}
\end{figure}

Utilizing Eq.~(\ref{eq22}) for numerical simulations with the dissipation of the phonon mode $\kappa_{ma}$ and the dephasing of NVs $\gamma_{NV}^1$ and $\gamma_{NV}^2$, the dynamic process can be described by the Lindblad master equation
\begin{equation}
  \begin{aligned}
    \dot{\hat{\rho}} = -i [\hat{H}_{RT}, \hat{\rho}] + \kappa_{ma}D(\hat{b}) \hat{\rho} \\
     + \gamma_{NV}^1 D(\hat{\sigma}_z^1) \hat{\rho} + \gamma_{NV}^2 D(\hat{\sigma}_z^2) \hat{\rho},
  \end{aligned}
\end{equation}
where $D(\hat{O})\hat{\rho}=\hat{O}\hat{\rho}\hat{O}^{\dagger}-\{\hat{O}^{\dagger}\hat{O}, \hat{\rho}\}/2$. As depicted in Figs.~\ref{fig7}(a)(b), when the fidelity of the phonon mode reaches its maximum value, a gate operation is finished. It reveals that 2-qubit gate time decreases as the squeezing parameter $r$ increases. Furthermore, Fig.~\ref{fig7}(c) indicates that with the squeezing parameter $r$ increasing, the gate time decreases dramatically. The 2-qubit-gate infidelity, arising from the dephasing of the spins is also affected by the squeezing parameter $r$ shown in Fig.~\ref{fig7}(d). It indicates that a 2-qubit gate with higher fidelity and shorter gate time can be achieved, with the fidelity being more than $99.9\%$ when the squeezing parameter $r=3$.

\section{\label{sec:V}Experimental feasibility}
To verify the experimental feasibility of the scheme proposed in this paper, we consider the scheme based on the experimental parameters given in ref.~\cite{2020GieselerP163604163604}, which include the radius $a=0.25\rm~{\mu}m$, cooling height $h_{cool}=3a$, equilibrium position $h_{eq}=3a$, cooling angle $\phi_{cool}=\pi/2$ and $\theta_{cool}=0$, equilibrium angle $\phi=\pi/2$ and $\theta=0$, the density of micromagnet $\rho=7430\rm~{kg/m^3}$, and residual induction $B_r=750\rm~{mT}$. In this case, the coupling strength between the NV center and the micromagnet is $2.9$~kHz, which is consistent with ref.~\cite{2020GieselerP163604163604}. A driving current is applied to the hybrid system, with the position $h_{cu}=2h_{eq}$ and amplitude $I_0=10\rm~{mA}$. The magnetic field induced by the current in the position of the NV center is $2\rm~{mT}$. If the static transverse magnetic field is $10$ times larger than the magnetic field induced by the driving current, the influence of the current on the NV center can be neglected.

\begin{figure}
  \centering
  \includegraphics[width=0.48\textwidth]{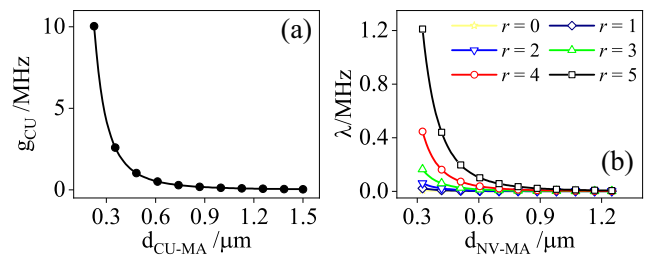}
  \caption{(a) the coupling strength between the driving current and the micromagnet varies with the distance between them. (b) indicates that the coupling strength of the NV and micromagnet grows as the distance between them decreases and the squeezing parameter $r$ increases.}
  \label{fig8}
\end{figure}

Fig.~\ref{fig8}(a) shows the coupling strength between the driving current and the micromagnet as a function of the distance between them $d_{CU-MA}$. As the distance between them grows, it decreases, reaching $10$~MHz when $d_{CU-MA}=0.3\rm~{\mu}m$. The coupling strength between the NV center and the micromagnet as a function of the distance between them $d_{NV-MA}$ and the squeezing parameter $r$ is depicted in Fig.~\ref{fig8}(b).  This shows that the coupling strength can be amplified when decreasing the distance and increasing the squeezing parameter $r$, reaching $1.2$~MHz if $d_{NV-MA}=0.3\rm~{\mu}m$ and the squeezing parameter $r=5$, indicating that it can reach the strong and even ultra-strong regime. To summarize, we can choose appropriate parameters based on the actual experimental conditions. In addition, the proposal is simple to implement under current experimental circumstances.

\section{\label{sec:VI}Conclusion}
Utilizing NV centers and a levitated micromagnet, we propose a hybrid quantum spin-mechanical system. A time-dependent driving current is applied to the hybrid system, which offers the critical nonlinear resource for the enhancement of the coupling strength. As a result, the spin-phonon and phonon-medicated spin-spin coupling strengths can be enhanced exponentially. The system can be utilized to construct an unconventional 2-qubit geometric phase gate with high fidelity and shorter gate time, as well as to prepare Schr\"odinger cat states with high fidelity. Furthermore, the Ground-state cooling approach, which requires the ultrastrong interaction between qubits and oscillators described by ref.~\cite{2021StreltsovP193602193602}, could be more simply implemented with this proposal. In addition, because the trapped frequency is related to the levitated height and the radius of the micromagnet, a wide frequency range can be easily obtained. Our proposal can also be extended to other solid-state spin systems, such as the silicon-vacancy center, germanium-vacancy center, and tin-vacancy center in diamond~\cite{2019BradacP56255625, 2020LiP153602153602, 2014HeppP3640536405}, allowing for more quantum information processing applications based on quantum levitodynamics.

\begin{acknowledgments}
This work was supported by the National Natural Science Foundation of China under Grant No. 92065105, and the Natural Science Basic Research Program of Shaanxi (Program No. 2020JC-02).
\end{acknowledgments}

\appendix
\section{\label{sec:appendixA}Levitation of the micromagnet}
In our scheme, we use a Type-II superconductor to levitate a micromagnet. The principle of levitation can be analytically analyzed by the frozen dipole model~\cite{1998KordyukP610612, 2020GieselerP163604163604}. As shown in Fig.~\ref{fig1}(c), the position vector of frozen and image dipoles are $\bm{R_f}=(-h_{cool},0,0)$ and $\bm{R_{\rm i}}=(-x,y,z)$, respectively. The orientation corresponds to $\bm{\mu_f}=\mu_m({\rm sin} \theta_{cool},0,-{\rm cos} \theta_{cool})$ and $\bm{\mu_{\rm i}}=\mu_m(-{\rm cos} \phi {\rm sin} \theta, {\rm sin} \phi {\rm sin} \theta, {\rm cos} \theta)$, respectively. The magnetic field produced by a dipole $\bm{\mu}$ at a position $\bm{r}$ is given by
\begin{equation}
  \bm{B} = \frac{\mu_0}{4 \pi} \left(\frac{3 \bm{r} \left(\bm{\mu \cdot r}\right)}{r^5}-\frac{\bm{\mu}}{r^3} \right).
  \label{eqA1}
\end{equation}
The effective magnetic field $\bm{B_{eff}}$ is composed of the magnetic field generated by the frozen dipole and image dipole, which is written as $\bm{B_{eff}}=\bm{B_f}+\bm{B_i}/2$ ~\cite{2020GieselerP163604163604}. In this case, the total potential energy of the levitated micromagnet is given by
\begin{equation}
  U = -\bm{\mu \cdot B_{eff}} + mgx,
  \label{eqA2}
\end{equation}
where $\bm{\mu}=\mu_m({\rm cos} \phi {\rm sin} \theta, {\rm sin} \phi {\rm sin} \theta, {\rm cos} \theta)$, $g=9.8m/s^2$, and $m$ is the mass of the micromagnet. For convenience, the potential energy in dimensionless form with $x_s=x/a$, $y_s=y/a$, and $z_s=z/a$ can be represented as
\begin{equation}
  U = U_s \left(\alpha_s x_s + g_u\right),
  \label{eqA3}
\end{equation}
where $\alpha_s=a/\alpha_{crit}$, $U_s=mg \alpha_{crit}$, $\alpha_{crit}=B_r^2/(16 g \rho \mu_0)$, and
\begin{subequations}
  \begin{align}
    &g_u = \frac{1+\left({\rm cos} \phi {\rm sin} \theta \right) ^ 2}{3x^3} \nonumber \\
    & - \frac{16}{3} \frac{g_c {\rm cos} \theta + g_s {\rm sin} \theta}{\left[\left(x + h_{cool}\right) ^ 2 + y^2 + z^2\right]^{5/2}} ,\\
    &g_c = \left[\left(x + h_{cool}\right) ^ 2 + y^2 -2 z^2\right] {\rm cos} \theta_{cool} \nonumber \\
    & + 3z \left(x + h_{cool}\right) {\rm  sin} \theta_{cool} ,\\
    &g_s = \left[-3z\left(x+h_{cool}\right) {\rm cos} \phi - 3zy {\rm sin} \phi \right] {\rm cos} \theta_{cool} \nonumber \\
    & +\Big{\{} \left[2 \left(x + h_{cool}\right)^2 - y^2 -z^2\right] {\rm cos} \phi \nonumber \\
    & + 3y\left(x + h_{cool}\right) {\rm sin} \phi \Big{\}} {\rm sin} \theta_{cool} .
  \end{align}
  \label{eqA4}
\end{subequations}
We now consider solely the potential energy along the $z$-direction, denoted by
\begin{equation}
  \mathrm{U}=U_{s}\left\{\frac{16}{3} \frac{2 z^{2}-\left(h_{e q}+h_{c o o l}\right)^{2}}{\left[\left(h_{e q}+h_{c o o l}\right)^{2}+z^{2}\right]^{5 / 2}}+\frac{1}{3 h_{e q}^{3}}+\alpha_{s} h_{e q}\right\},
  \label{eqA5}
\end{equation}
where $h_{eq}$ and $h_{cool}$ correspond to the equilibrium position and the cooling height, respectively. According to the analysis in part~\ref{sec:IIB}, we assign $\phi=\phi_{cool}=\pi/2$ and $\theta=\theta_{cool}=0$, which means that the direction of levitation represented by $\theta$ and $\phi$ is the same as the initial orientation $\theta_{cool}$ and $\phi_{cool}$. This indicates that the rotation of the micromagnet is neglected. By expanding at the equilibrium position and removing the constant and high order components, the potential energy can be written as simple harmonic potential
\begin{equation}
  U = \frac{1}{2} k_{ma} z^2 ,
  \label{eqA6}
\end{equation}
where
\begin{equation}
  k_{ma} = \frac{\mu_0 \mu_m^2}{4 \pi} \frac{3}{\left(h_{eq} + h_{cool}\right)^5}.
  \label{eqA7}
\end{equation}
The motion of the levitated micromagnet along the $z$-direction can be regarded as a simple harmonic motion, as represented by
\begin{equation}
  \hat{H}_{magnet} = \frac{\hat{p}^2}{2m} +\frac{1}{2} k_{ma} \hat{z}^2 ,
  \label{eqA8}
\end{equation}
where $\hat{p}_z$ and $\hat{z}$ are momentum and position operators, respectively. By quantizing the Hamiltonian with $\hat{p}_z=-i\sqrt{m\omega_{ma}/2} (\hat{a}-\hat{a}^{\dagger})$, $\hat{z}=z_0 (\hat{a} + \hat{a}^{\dagger})$, and $z_0=\sqrt{1/ (2m \omega_{ma})}$, we can obtain
\begin{equation}
  \hat{H}_{magnet} = \omega_{ma} \hat{a}^{\dagger} \hat{a},
  \label{eqA9}
\end{equation}
where $\omega_{ma}=\sqrt{k_{ma}/m}$ represents the trapping frequency associated with the cooling conditions, and $\hat{a}$ ($\hat{a}^{\dagger}$) represents the annihilation (creation) operator.
\section{\label{sec:appendixB}Interaction between NV and micromagnet}
The NV center is coupled to the levitated micromagnet via a strong magnetic field gradient. Firstly, we investigate how one NV center interacts with a micromagnet. In our proposal, the micromagnet is described as a dipole with $\bm{\mu}=\mu_m({\rm cos} \phi {\rm sin} \theta, {\rm sin} \phi {\rm sin} \theta, {\rm cos} \theta)$ at position $(x, y,z)$. At position $(h_{eq},0,d)$ of the NV center, the magnetic field induced by the micromagnet along the $z$-direction is given by
\begin{equation}
  \bm{B_{ma}} = \frac{\mu_0 \mu_m}{2 \pi \left(d-z\right)^3} \bm{\hat{e}_z}.
  \label{eqB1}
\end{equation}
By expanding at the equilibrium position, removing the constant and high order terms, the magnetic field can be written as
\begin{equation}
  \bm{B_{ma}} = \frac{3 \mu_0 \mu_m}{4 \pi d^4} z \bm{\hat{e}_z}.
  \label{eqB2}
\end{equation}
Therefore, the interaction Hamiltonian is given by
\begin{equation}
  \hat{H}_{int} = \gamma_e \frac{3 \mu_0 \mu_m}{4 \pi d^4} \hat{z} \hat{S}_z.
  \label{eqB3}
\end{equation}
 Quantizing the Hamiltonian, we can obtain
\begin{equation}
  \hat{H}_{int} = \lambda \left(\hat{a} + \hat{a}^{\dagger} \right) \hat{S}_z,
  \label{eqB4}
\end{equation}
where $\lambda=\gamma_e B_r a ^3 z_0 / d^4$ is the coupling strength.

What is more intriguing is that two NVs are symmetrically placed at positions $(h_{eq},0,d)$ and  $(h_{eq},0,-d)$ on either side of the micromagnet along the direction of magnetization. Along the $z$-axis, the magnetic field produced by the micromagnet is given by
\begin{subequations}
  \begin{align}
    &\bm{B_1} = \frac{\mu_0 \mu_m}{2 \pi \left(d-z\right)^3} \bm{\hat{e}_z},\\
    &\bm{B_2} = \frac{\mu_0 \mu_m}{2 \pi \left(d+z\right)^3} \bm{\hat{e}_z}.
  \end{align}
  \label{eqB5}
\end{subequations}
After expanding at the equilibrium position, omitting constant items and high order components, the magnetic field can be represented as
\begin{subequations}
  \begin{align}
    &\bm{B_1} = \frac{3 \mu_0 \mu_m}{4 \pi d^4} z \bm{\hat{e}_z}, \\
    &\bm{B_2} = -\frac{3 \mu_0 \mu_m}{4 \pi d^4} z \bm{\hat{e}_z}.
  \end{align}
  \label{eqB6}
\end{subequations}
In the same way as the one-NV process, we can get the interaction Hamiltonian of two NV centers, which reads
\begin{equation}
  \hat{H}_{int} = \lambda \left(\hat{a} + \hat{a}^{\dagger}\right) \left(\hat{S}_z^1 - \hat{S}_z^2\right),
  \label{eqB7}
\end{equation}
where $\lambda=\gamma_e B_r a ^3 z_0 / d^4$ is the coupling strength.
\section{\label{sec:appendixC}The Hamiltonian of drive current}
Current pumping $I_{cu}(t)=I_0 {\rm cos} 2\omega_{cu}t$ is added to the hybrid system to enhance coupling strength. The position of origin current and image current in the $zx$-plane are $\bm{R_{or}}=(h_{cu},0,0)$ and $\bm{R_{im}}=(-h_{cu},0,0)$, respectively. When just the magnetic field in the $z$-direction near the equilibrium position is considered, the total magnetic field created by the origin and image current is given by
\begin{equation}
  \bm{B_{cu}} = \bm{B_{or}} + \bm{B_{im}} = \frac{\mu_0 I}{2 \pi} \left(\frac{1}{r_{or}} + \frac{1}{r_{im}}\right)\bm{\hat{e}_z}.
  \label{eqB8}
\end{equation}
Then the potential energy of the micromagnet in the magnetic field generated by the current can be written as
\begin{subequations}
  \begin{align}
    &U_{cu} = -\bm{\mu \cdot B_{cu}} , \\
    &U_{c u}=-\frac{\mu_{0} \mu_{m} I}{2 \pi}\left[\frac{1}{\sqrt{\left(h_{e q} - h_{c u}\right)^{2}+z^{2}}}\right] \nonumber \\
    &-\frac{\mu_{0} \mu_{m} I}{2 \pi}\left[\frac{1}{\sqrt{\left(h_{e q}+h_{c u}\right)^{2}+z^{2}}}\right] .
  \end{align}
  \label{eqC1}
\end{subequations}
By expanding at the equilibrium position, dropping constant items and high order terms, the magnetic field can be represented as
\begin{equation}
  U_{cu} = - \frac{1}{2} k_{cu} \hat{z}^2 {\rm cos} 2 \omega_{cu} t ,
  \label{eqC2}
\end{equation}
where
\begin{equation}
  k_{cu} = \frac{\mu_0 \mu_m I_0}{2 \pi} \left[\frac{1}{\left(h_{cu} - h_{eq}\right)^3} +\frac{1}{\left(h_{cu} + h_{eq}\right)^3}\right] .
  \label{eqC3}
\end{equation}
Quantizing the potential energy, the Hamiltonian is given by
\begin{equation}
  \hat{H}_{cu} = -g_{cu}\left(\hat{a} + \hat{a}^{\dagger} \right) ^2 {\rm cos} 2\omega_{cu} t,
  \label{eqC4}
\end{equation}
where $g_{cu}=k_{cu} z_0 ^2 / 2$ is the coupling strength between the drive current and the micromagnet.

\section{\label{sec:appendixD}The total Hamiltonian of the hybrid system (two NVs)}
This section will derive the Hamiltonian, which describes the interaction between two NVs and the micromagnet. Two NV centers are symmetrically arranged on either side of the micromagnet, coupling to the center of mass motion of the micromagnet through a strong magnetic field gradient produced by the micromagnet. The Hamiltonian of the hybrid system is given by
\begin{equation}
  \begin{aligned}
    \hat{H}_{TT}=\frac{\omega_{0}}{2} \left(\hat{\sigma}_{z}^1 + \hat{\sigma}_{z}^2\right) + \omega_{m a} \hat{a}^{\dagger} \hat{a} \\ + \Lambda \left(\hat{a} + \hat{a}^{\dagger}\right)\left(\hat{\sigma}_{x}^{1}-\hat{\sigma}_{x}^{2}\right)
    -g_{c u}\left(\hat{a} + \hat{a}^{\dagger}\right)^{2} {\rm cos} 2 \omega_{c u} t .
  \end{aligned}
  \label{TwoNvHT}
\end{equation}
Moving in the rotation frame, the Hamiltonian can be simplified as
\begin{equation}
  \begin{aligned}
    \hat{H}_{TT}=\frac{\omega_{0}}{2} \left(\hat{\sigma}_{z}^1 + \hat{\sigma}_{z}^2\right) + \omega_{ma} \hat{a}^{\dagger} \hat{a} \\
    + \Lambda \left[\hat{a}\left(\hat{\sigma}_{1}^{+}-\hat{\sigma}_{2}^{+}\right) + \hat{a}^{\dagger} \left(\hat{\sigma}_{1}^{-}-\hat{\sigma}_{2}^{-}\right)\right]
    -\frac{g_{c u}}{2}\left(\hat{a}^{2} + \hat{a}^{\dagger^{2}}\right) .
  \end{aligned}
\end{equation}
Utilizing the Bogoliubov transformation~\cite{2016LemondeP1133811338, 2019BurdP11631165, 2021BurdP898902} $\hat{b}=\hat{a}^{\dagger} {\rm cosh} r - \hat{a} {\rm sinh} r$ and ${\rm tanh} 2r=g_{cu}/\delta_m$, the total
Hamiltonian can be diagonalized and represented as
\begin{equation}
  \begin{aligned}
    H_{RT}=\frac{\delta_{0}}{2}\left(\hat{\sigma}_{z}^{1}+\hat{\sigma}_{z}^{2}\right)+\Delta_{m} \hat{b}^{\dagger} \hat{b} \\
    +\Lambda_{e f f}\left(\hat{b} + \hat{b}^{\dagger}\right)\left(\hat{\sigma}_{x}^{1}-\hat{\sigma}_{x}^{2}\right) .
  \end{aligned}
\end{equation}
Without loss of generality, we assume the $\delta_0=0$, such that the Hamiltonian can be simplified by a Schrieffer-Wolff transformation~\cite{2004WilsonRaeP7550775507, 2013AlbrechtP8301483014} $\hat{H}_{RT}^{eff}=e^S\hat{H}_{RT}e^{-S}$, where $S=\eta (\hat{b}^{\dagger}-\hat{b})  (\hat{\sigma}_x^1 - \hat{\sigma}_x^2)$ with $\eta=\Lambda_{eff}/\Delta_m$ and the Lamb-Dicke condition $\eta \ll 1$. The effective Hamiltonian is given by
\begin{equation}
  \hat{H}_{RT}^{eff} = \Delta_m \hat{b}^{\dagger} \hat{b} - \xi \left(\hat{b} + \hat{b} ^ {\dagger}\right) \left(\hat{\sigma}_x^1-\hat{\sigma}_x^2\right)^2 ,
\end{equation}
where $\xi = \Lambda_{eff}^2/\Delta_m$. Retaining only the terms containing $\xi$, we obtain the Ising interaction Hamiltonian
\begin{equation}
  \hat{H}_{Ising} = \xi \left(\hat{\sigma}_x^1 - \hat{\sigma}_x^2\right)^2 ,
\end{equation}
which corresponds to the one-axis twisting interaction.

%\bibliography{ArticleReference}

\end{document}